\documentclass{aa}

\usepackage{graphicx}
\usepackage{txfonts}
\usepackage{color}
\usepackage{xspace}
\usepackage{siunitx}
\usepackage{amsmath}
\usepackage{amssymb}
\usepackage{xcolor}
\usepackage{dashbox}
\usepackage{framed}
\usepackage{lipsum}
\usepackage{placeins}
\usepackage{stfloats}

\DeclareSIUnit{\mas}{mas}
\DeclareSIUnit{\arcsec}{arcsec}
\DeclareSIUnit{\arcmin}{arcmin}
\DeclareSIUnit{\degree}{deg}
\DeclareSIUnit{\year}{yr}
\DeclareSIUnit{\msun}{M_{\odot}}
\DeclareSIUnit{\mjup}{M_{\mathrm{Jup}}}
\DeclareSIUnit{\pixel}{pix}

\usepackage{hyperref}

\hypersetup{
        colorlinks=true, 
        breaklinks=true,
        linkcolor=blue, 
        citecolor=blue, 
        filecolor=blue, 
        urlcolor=blue,
        unicode=false, 
        pdftoolbar=true, 
        pdfmenubar=true, 
        pdffitwindow=false, 
        pdfstartview={Fit}, 
        pdftitle={VISIONS - The VISTA Star Formation Atlas}, 
        pdfauthor={Stefan Meingast}, 
        pdfsubject={II. The data processing pipeline},
        pdfcreator={Stefan Meingast}, 
        pdfkeywords={},
        pdfnewwindow=true, 
        pdfdisplaydoctitle=true 
}

\makeatletter
\renewcommand*\aa@pageof{, page \thepage{} of \pageref*{LastPage}}
\makeatother

\begin{document}

\title{VISIONS: The VISTA Star Formation Atlas}
\subtitle{II. The data processing pipeline\thanks{Based on observations collected at the European Organization for Astronomical Research in the Southern Hemisphere under ESO programs 198.C-2009 and 090.C-0797(A).}}

\author{
        Stefan Meingast\inst{1} \and
        Hervé Bouy\inst{2} \and
        Verena Fürnkranz \inst{3} \and
        David Hernandez \inst{1} \and
        Alena Rottensteiner\inst{1} \and
        Erik Brändli\inst{1}
        }

\institute{Universität Wien, Institut für Astrophysik, T\"urkenschanzstrasse 17, 1180 Wien, Austria 
\\ \email{stefan.meingast@univie.ac.at}
\and
Laboratoire d’Astrophysique de Bordeaux, Univ. Bordeaux, CNRS, B18N, Allée Geoffroy Saint-Hillaire, 33615 Pessac, France
\and
Max-Planck-Institut für Astronomie, Königstuhl 17, D-69117, Heidelberg, Germany
}

\date{Received 22 December 2022  / accepted 28 February 2023}

\abstract{The VISIONS public survey provides large-scale, multi-epoch imaging of five nearby star-forming regions at sub-arcsecond resolution in the near-infrared. All data collected within the program and provided by the European Southern Observatory (ESO) science archive are processed with a custom end-to-end pipeline infrastructure to provide science-ready images and source catalogs. The data reduction environment has been specifically developed for the purpose of mitigating several shortcomings of the bona fide data products processed with software provided by the Cambridge Astronomical Survey Unit (CASU), such as spatially variable astrometric and photometric biases of up to \SI{100}{mas} and \SI{0.1}{mag}, respectively. At the same time, the resolution of co-added images is up to \SI{20}{\percent} higher compared to the same products from the CASU processing environment. Most pipeline modules are written in Python and make extensive use of C extension libraries for numeric computations, thereby simultaneously providing accessibility, robustness, and high performance. The astrometric calibration is performed relative to the \textit{Gaia} reference frame, and fluxes are calibrated with respect to the source magnitudes provided in the Two Micron All Sky Survey~(2MASS). For bright sources, absolute astrometric errors are typically on the order of 10 to \SI{15}{mas} and fluxes are determined with sub-percent precision. Moreover, the calibration with respect to 2MASS photometry is largely free of color terms. The pipeline produces data that are compliant with the ESO Phase 3 regulations and furthermore provides curated source catalogs that are structured similarly to those provided by the 2MASS survey.}



\maketitle

\section{Introduction}
\label{sec:introduction}

VISIONS is a public survey carried out at the European Southern Observatory (ESO) with the Visible and Infrared Survey Telescope for Astronomy \citep[VISTA;][]{vista} and the VISTA Infrared Camera \citep[VIRCAM;][]{vircam}. The survey has actively collected data from April 2017 to March 2022, providing images of five nearby star-forming regions. In total, more than \SI{70000} exposures (pawprints) have been recorded that cover an area of about \SI{650}{deg^2} in the near-infrared bands $J$, $H$, and $K_S$. A detailed description of the survey, regarding scientific motivation, the covered regions, the observing strategy, and the data contents, is available in the first paper in this series \citep{VISIONSI}.

VISTA typically acquires a data volume of more than \SI{200}{GB\per night}. Combined with the complex nature of near-infrared imaging, this large volume calls for a dedicated data-processing infrastructure to ensure an efficient utilization of the telescope time and to maximize the scientific return. Most public surveys rely on the processing facilities at the Cambridge Astronomical Survey Unit (CASU\footnote{\href{http://casu.ast.cam.ac.uk}{http://casu.ast.cam.ac.uk}}). However, as previously discussed by \citet{meingast16}, their published data-processing recipes can lead to sub-optimal results regarding photometry, astrometry, and image quality. To mitigate these issues, \citet{meingast16} developed a processing workflow that is specifically geared to their collected data. The findings and core algorithms of this work constitute the basis of the VISIONS pipeline.

Due to the large data volume for VISIONS, we have further developed the implemented algorithms into a pipeline infrastructure. This data-processing environment, called \texttt{vircampype}, operates independently from CASU, and is available as a public GitHub repository\footnote{\href{https://github.com/smeingast/vircampype}{https://github.com/smeingast/vircampype}}. In this paper, we present a detailed description of the data flow and the implemented modules in Sect.~\ref{sec:pipeline}, and evaluate the performance by comparing our results to publicly available data products processed with the CASU workflow~(Sect.~\ref{sec:performance}).

\section{Pipeline description}
\label{sec:pipeline}

\begin{figure*}
        \centering
    \resizebox{\hsize}{!}{\includegraphics[]{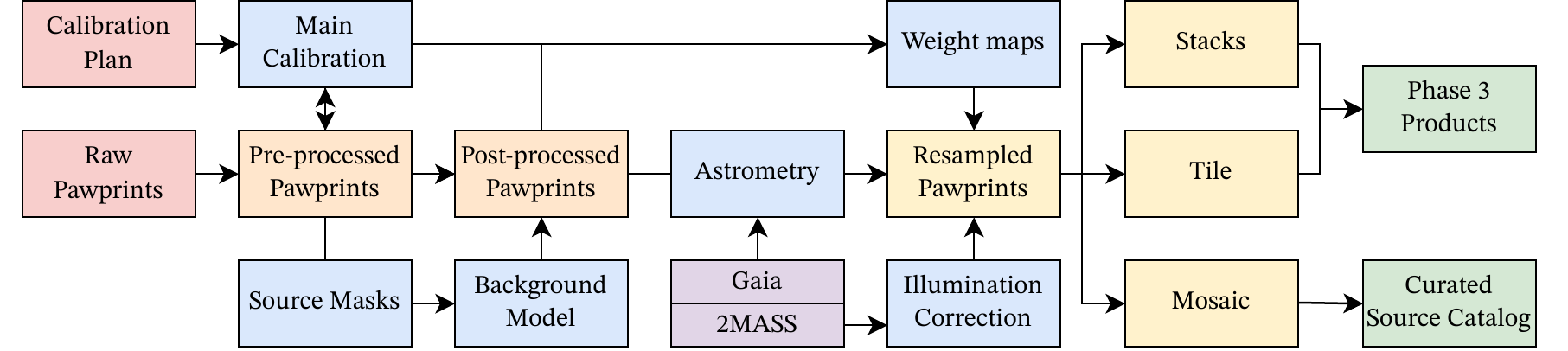}}
        \caption{Pipeline overview chart depicting a simplified version of the data flow within the VISIONS pipeline. Each tile represents a specific processing stage where one or more recipes are executed. Tiles with a red background indicate raw input data obtained from the ESO science archive. Blue tiles denote calibration files, such as flat fields and tables to linearize the detector response. Orange tiles refer to intermediate science data products. Yellow tiles mark processed science data from which publication-ready products (green tiles) are generated. The purple tile indicates data from external sources, which correspond to the source catalogs from the 2MASS survey and the \textit{Gaia} mission. The arrows mark the interdependencies between the consecutive processing stages.}
        \label{img:flowchart}
\end{figure*}

In this section, we present an overview of the data-processing pipeline used for the VISIONS survey. The data flow in the pipeline starts with the creation of the main calibration files that are required to remove the instrumental signature from VIRCAM raw data. Given these calibration files, the pipeline deploys a series of consecutive data-processing recipes to produce science-ready images and source catalogs. Figure~\ref{img:flowchart} presents a simplified visualization of the data flow within the system. In this figure, independent processing steps are given as tiles that are connected with arrows specifying their interdependencies. The colors of the tiles discriminate between different types of data, indicating raw data in red, calibration files in blue, intermediate science data in orange, processed science data in yellow, and publication-ready products in green. Following the indicated data flow, the subsequent sections describe each individual step in detail.

For several stages of the pipeline, Astromatic software packages\footnote{\href{https://www.astromatic.net/software/}{https://www.astromatic.net/software/}} are used, including SExtractor \citep{sextractor}, Scamp \citep{scamp}, and SWarp \citep{swarp}. To run the pipeline, it is required that these tools be preinstalled, with binaries being reachable through the system shell. The pipeline includes built-in wrappers that allow for them to be called from within the recipe execution sequence in Python for all Astromatic tools used.

\subsection{Main calibration files}
\label{ssec:main_calibration}

For the VISIONS program, no custom observation sequences have been requested and all required calibration files have been recorded as part of the VIRCAM calibration plan\footnote{\href{www.eso.org/sci/facilities/paranal/instruments/vircam/doc.html}{www.eso.org/sci/facilities/paranal/instruments/vircam/doc.html}}. The main calibration files generated by the pipeline are used to remove the instrumental signature from the raw science data and include bad pixel masks, polynomial coefficients for correcting detector nonlinearity, measurements of the dark current, read noise and gain, as well as flat fields.

The pipeline recipes that generate the required calibration files work independently from all other pipeline modules and are executed prior to processing science data. This is indicated in the top left corner of Fig.~\ref{img:flowchart}, where the pipeline uses raw data from the calibration plan to generate the main calibration files. Given an arbitrary number of VIRCAM files that are produced as part of the calibration plan, the pipeline categorizes all input files to separate different image types and to group them according to their recording time stamps. For each group, e.g. dark frames with unique combinations of detector integration times (DIT) and number of DIT (NDIT) exposures, the pipeline consecutively executes several recipes to build a main calibration library. The resulting calibration files are later matched with the science frames for data processing.

\subsubsection{Bad pixel mask}
\label{sssec:bpm}

The VISIONS pipeline uses masks to identify individual bad pixels. These masks are important for determining robust image statistics, such as the detectors' read noise and gain, as well as the flux component contributed by the sky background. Moreover, bad pixels also affect their immediate neighbors during the resampling stage, where all science frames are reprojected onto a common reference grid. If not properly masked, individual bad pixels lead to entire clusters of incorrectly resampled and thereby unusable pixel values in the processed data products. 

In reference to the VISIONS pipeline, bad pixel masks are built from a dedicated calibration sequence in the VIRCAM calibration plan that features a series of dome flats with constant exposure time. From these flats, initially, a raw dark frame is subtracted. Next, the median flux levels in each frame are used to normalize the entire exposure sequence so that all pixel values are close to unity. Subsequently, all frames for a given detector are stacked and pixels are flagged according to their deviation from the median value in each pixel column. The deviation allowed can be configured in the pipeline setup and is set to \SI{4}{\percent} by default. For each pixel column, the pipeline recipe counts the number of flagged pixels and determines bad pixels as those that are marked in at least (by default) \SI{20}{\percent} of the images in the flat-field sequence. In this way, the pipeline identifies pixels that show an irregular response under constant illumination. In addition, the pipeline collapses the stacked flat fields for each detector by calculating the median value for each pixel column, excluding previously identified bad values. Any pixels with values close to 0 in the combined image are also added to the bad pixel mask. As a quality control parameter, the pipeline counts the fraction of bad pixels for each detector. Typically between 0.2 and \SI{0.5}{\percent} are identified as bad pixels; only detectors 1 and 16 feature values above \SI{1}{\percent}.

\begin{figure}
        \centering
        \resizebox{\hsize}{!}{\includegraphics[]{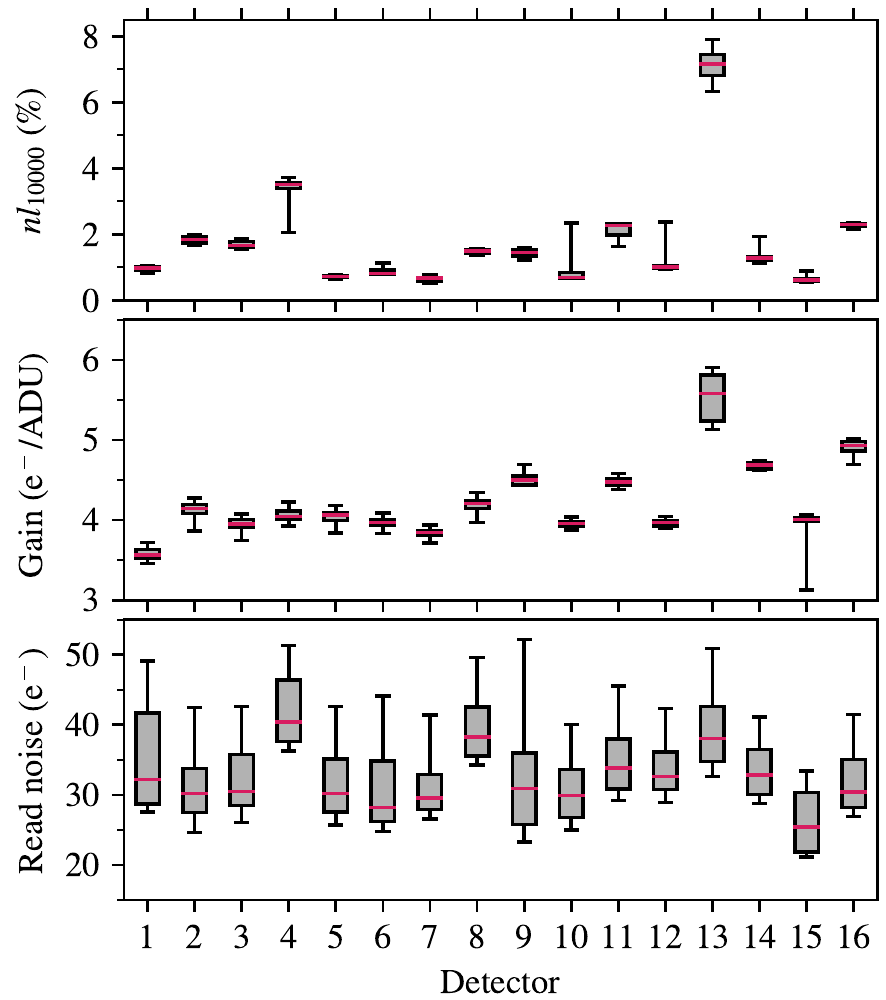}}
        \caption{Box plot of detector characteristics as determined from the main calibration files that were built from observations obtained during the VISIONS program. In all panels, the red line marks the median value, the boxes indicate the interquartile range, and the whiskers represent the fifth and 95th percentile. The panel at the top displays the measured nonlinearity at \SI{10000}{ADU} ($nl_{10000}$) for each detector. The panels in the center and at the bottom show the gain and read noise, respectively.}
        \label{img:gain_rdnoise_linearity}
\end{figure}

\subsubsection{Nonlinearity}
\label{sssec:nonlinearity}

To correct for nonlinearity in the response of the detectors, we have adopted the same procedure as described in the VISTA Data Reduction Library Design document available through the CASU online resources\footnote{\href{http://casu.ast.cam.ac.uk}{http://casu.ast.cam.ac.uk}}. To summarize briefly, the VIRCAM data acquisition system provides only reset-corrected frames from the double-correlated sampling (DCS) readout mode. As a consequence, only the difference between two nonlinear readouts $\Delta I' = I_2' - I_1'$ is available, where $I_1'$ and $I_2'$ are the first (reset frame) and second readouts, respectively. To transform the measured nonlinear values $\Delta I'$ to their linear counterparts $\Delta I$, a polynomial model was constructed. It takes the form
\begin{align}
    \Delta I' &= \sum_m a_m \Delta I^m \left[ \left( 1 + k \right)^m - k^m \right] \label{equ:linearity1} \\
    \Delta I' &= \sum_m b_m t^m \left[ \left( 1 + k \right)^m - k^m \right] \label{equ:linearity2},
\end{align}
where the exposure time $t$ in Equ.~\ref{equ:linearity2} substitutes $\Delta I$ in Equ.~\ref{equ:linearity1} via $\Delta I = s \cdot t$ and $b_m = a_m \cdot s^m$, with $s$ being an unknown scaling factor that relates the exposure time to the linear DCS readout. The variable $k$ is related to the reset-read overhead and is given by the ratio of the minimum exposure time $t_\mathrm{min}=1.0011\,\si{s}$ to the total exposure time of any given image. The order $m$ of the polynomial was set to a fixed value of three for the case of processing data recorded in the VISIONS program. Equation~\ref{equ:linearity2} allowed us to solve for coefficients $b_m$, given a specifically designed dome flat sequence in the VIRCAM calibration that features increasing exposure times at a constant illumination level. Given the measured flux levels in the calibration sequence, and requiring $a_1 = 1$, by definition $a_m = b_m / b_1^m$ can be used to obtain the coefficients $a_m$ required to compute the linearized pixel values. For each dome flat sequence associated with the VISIONS observations, both $a_m$ and $b_m$ have been saved in tabular format for further processing of the raw data.

Given the coefficients $a_m$, linearized values can be obtained for each pixel individually by computing the roots of Equ.~\ref{equ:linearity1}. To this end, the pipeline implements the method developed by Henry Freudenreich for the IDL Astronomy User's Library\footnote{\href{https://idlastro.gsfc.nasa.gov}{https://idlastro.gsfc.nasa.gov}}. Due to vectorization and the use of C extension libraries, the implemented inversion method processes about \SI{100}{Mpix/s} with the available hardware for testing (Apple M1 Pro), that is to say the pipeline requires less than one second to linearize an entire VIRCAM pawprint data cube.

For quality control purposes, the pipeline calculates the ratio of linearized to nonlinear values for a nonlinear input pixel value of \SI{10000}{ADU}, $nl_{10000}$. In comparison, the CASU pipeline environment computes a similar metric that measures the nonlinearity as the ratio of the slope of the polynomial fit at \SI{10000}{ADU} to its linear component (i.e., $b_1$ in Equ.~\ref{equ:linearity2}). As such, the quality control parameters that measure the nonlinearity in the VISIONS and CASU pipelines yield different values and they are not directly comparable. The top panel of Fig.~\ref{img:gain_rdnoise_linearity} displays a box plot for all measured nonlinearity values of a given detector over the duration of the VISIONS program. Typically, $nl_{10000}$ takes values around 1 and \SI{2}{\percent}; only detector 13 exhibits a significantly larger nonlinearity with about \SI{7}{\percent} at \SI{10000}{ADU}. In addition, the figure reveals that the detector nonlinearity was relatively stable for the duration of the VISIONS survey; only a few detectors display individual outliers.

\subsubsection{Dark current}
\label{sssec:dark_current}

The main dark frames have been built for each unique DIT-NDIT combination. Each frame in a sequence is first scaled so that the pixel values correspond to $\mathrm{NDIT}=1$. Subsequently, the pipeline linearizes all pixel values, as described in Sect.~\ref{sssec:nonlinearity}, and stacks all frames for an individual detector into data cubes. At this stage, the pipeline recipe optionally masks the minimum and maximum values for each stacked pixel column. The main dark calibration file is obtained by calculating an arithmetic mean of all stacked frames\footnote{For the typical dark files obtained with the VIRCAM calibration plan, we preferred to compute the mean (as opposed to the median) because of the low number of supplied images.}. For the dark frames, typically only five pawprints are collected for each DIT-NDIT combination.

The pipeline calculates the mean dark current in ADU/s for each detector since this step precedes the computation of the gain in the workflow. Typical values for most detectors are \SI{0.1}{ADU/s}, with detectors 1 through 4 having a larger, but still small, dark current of about \SI{0.4}{ADU/s}. For detector 13, we measured the largest dark current, corresponding to roughly \SI{1.5}{ADU/s}. Considering these values, the dark current represents a negligible contributor to the total noise budget for typical VISIONS observations.

\subsubsection{Gain and read noise}
\label{sssec:gain_rdnoise}

The detector gain and read noise are computed from a custom observation template, as defined in the VIRCAM calibration plan. The input consists of two dome flats and their corresponding dark frames that allow one to compute both values at the same time via
\begin{align}
    \mathrm{Gain}\,(\mathrm{e}^-/\mathrm{ADU}) &= \frac{(\bar{F}_1 + \bar{F}_2) - (\bar{D}_1 + \bar{D}_2)}{\sigma^2_{F_1 - F_2} - \sigma^2_{D_1 - D_2}} \\
    \mathrm{Read\,noise}\,(\mathrm{e}^-) &= \frac{\mathrm{Gain} \times \sigma_{D_1 - D_2}}{\sqrt{2}},
\end{align}
where $\bar{F}_1$, $\bar{F}_2$, $\bar{D}_1$, and $\bar{D}_2$ are the mean values of the flat and dark frames, respectively, and $\sigma^2_{F_1 - F_2}$ and $\sigma^2_{D_1 - D_2}$ are the variances of the difference images constructed from the input files \citep{Howell06}. Both gain and read noise were computed at the detector level, and Fig.~\ref{img:gain_rdnoise_linearity} displays statistics for all values obtained throughout the duration of the VISIONS survey. For the gain, we measured stable values with a few outliers. Similarly, the read noise usually takes values between 30 and \SI{40}{e^-}. We note here that these statistics were obtained separately for each readout channel in the CASU pipeline. However, we did not measure significant differences within individual detectors, and therefore we have obtained these characteristics for each detector as a whole. The obtained values for the detector gain and read noise have only been used in subsequent pipeline recipes to compute noise properties correctly for source extraction and cosmic ray detection, for example. In particular we note here that the calculated parameters cannot be used as scaling factors to harmonize the gain across the detector array due to the nonuniform illumination of the dome flat screen.

\subsubsection{Flat-field}
\label{sssec:flat_field}

As part of the VIRCAM calibration plan, twilight flats in each filter are taken on a regular basis. For these frames, the pipeline initially applies a bad pixel mask, linearizes the pixel values, and subtracts the dark current, using the main calibration files as detailed above. After these initial steps, all twilight flats are normalized by their respective mean flux value, so that each image features values close to unity. Subsequently, all frames collected for a given detector are stacked. By default, the flat-field recipe masks the minimum and maximum pixels, pixels that deviate by more than \SI{70}{\percent} from unity, and all pixels outside a 3-$\sigma$ range around the median for each pixel column in the stack. Next, the mean fluxes measured prior to normalization are used to calculate a weighted average that combines all frames for each detector into a main flat field. Hence, flat-field data with more flux, that is to say with a better signal-to-noise, receive a larger weight compared to images with low illumination levels. At this stage, the mean values of the flat fields for each detector are close to unity because of the preceding normalization sequence. As a consequence, no information on the relative differences between the received signals of the detectors (e.g., due to vignetting) is taken into account. To restore the relative flux differences between the detectors, the initially measured fluxes are used as scaling factors.

For quality control and data-processing purposes, the pipeline saves all measured mean fluxes and scaling factors between the detectors. Moreover, the created main flat fields serve as an initial global weight for the creation of individual weight maps at a later stage in the data flow. This dependency is visualized in Fig.~\ref{img:flowchart}, where both sets of data from the main calibration module and processed pawprints are used to create individual weight maps (see Sect.~\ref{ssec:weightmaps}).

\subsection{Pawprint pre-processing}
\label{ssec:pre-processing}

In order to provide optimized background models (see Fig.~\ref{img:flowchart}), we have deliberately introduced separate pre-processing and post-processing stages. During the pre-processing step, all products in the main calibration library are applied to the raw science pawprints. In addition, several entries to the image headers are added or modified.

Pre-processing starts with dividing all raw science pawprints by their respective NDIT values, followed by removing the nonlinearity as outlined in Sect.~\ref{sssec:nonlinearity}. Next, a main dark frame with matching DIT and NDIT values is subtracted from each pawprint, and the flat field is applied to remove pixel-to-pixel variations in the system response, for example due to vignetting and a variable gain.

Following the initial processing step, the pipeline resets the world coordinate information in the FITS headers of the files. This step is necessary since, for some files, the telescope pointing information (CRVAL1 and CRVAL2) on the observed field can be missing. Information on the approximate pointing is instead taken from the telescope target coordinates, which we have found to be reliably present in all image files. Furthermore, by default, the translation of pixel-to-world coordinates in VIRCAM images is described by a zenithal polynomial projection for the entire detector array. We have found that this default projection is largely incompatible with software that is used later during processing. For this reason, the projection information in the headers has been converted to a gnomonic tangential form at the detector level, while simultaneously resetting all information on the detector arrangement in the focal-plane array. For this step, a world coordinate system was fitted to the uncalibrated footprint of each image, replacing the default projection parameters. Moreover, several data properties have been set in the FITS headers, including gain, read noise, saturation levels, default atmospheric extinction values, and the airmass at the time of the observation. Where necessary, these values have been scaled to match the normalization that occurred in the previous data manipulation stages. The resulting data products of the preceding steps are stored as an intermediate processing step.

\subsection{Pawprint post-processing}
\label{ssec:post-processing}

The post-processing stage applies a series of modifications aimed at removing the remaining instrumental signatures from the pre-processed pawprints. Specifically, the pipeline executes recipes that create source masks as well as background models, and optionally runs a series of cosmetic modifications.

Following the pipeline data flow in Fig.~\ref{img:flowchart}, the pre-processed pawprints are used to create source masks. The purpose of source masks is to enable a robust determination of background statistics and the construction of a reliable background model (bad pixel masks are already applied during flat-fielding). In the first step, the pipeline creates a preliminary background model. To this end, the recipe calculates an initial estimate of the background using iterative clipping, similar to the DAOPHOT MMM algorithm \citep{daophot}, and masks all pixels with values three sigma above this threshold. Subsequently, the pipeline scales all data to match the mean background level and computes the preliminary background model as the median of the clipped and scaled image stack.

The preliminary background model is subtracted from the processed frames, and source masks are created by determining pixels that exceed a user-defined threshold. By default, sources are identified as clusters of pixels with a minimum of three adjacent pixels with values that are three sigma levels above the local background value. Optionally, a maximum size of a pixel cluster can be defined (by default \SI{25000}) and bright sources from the Two Micron All Sky Survey~(2MASS) catalog can also be automatically masked with a circular patch. In addition, it is possible to manually specify regions in DS9 format (e.g., extended emission) that act as additional masks. However, masking bright 2MASS sources and manually specified regions can be inaccurate because the astrometric information in the FITS headers at this stage is typically only correct to within an arcminute.

Following the creation of source masks, a final background model is computed similarly to the static background model described above. The main differences are the application of source masks prior to merging the frame stacks and the possibility of setting a time window for creating multiple background models over the duration of the observation block. By default, the pipeline groups all data collected within \SI{1}{\hour}, which, for the observation blocks in the VISIONS program, groups all images into a single set. Setting this value to shorter time intervals creates a series of background models, with the disadvantage that fewer images are used, thereby introducing more noise into each background model. 

Finally, the background model is subtracted from the pre-processed images. At this stage, the pipeline optionally interpolates bad pixels as given in the bad pixel mask using a Gaussian-shaped kernel with a standard deviation of one pixel. This option is activated by default, but only interpolates isolated bad pixels, in our case referring to entries in the bad pixel mask that form a cluster with a maximum size of three connected pixels. Deactivating this step results in retaining a large number of bad pixels that are later propagated into the resampled pawprints in the form of clusters of bad pixels. Moreover, a recipe for correcting a horizontally striped pattern that is picked up in each detector during readout is executed. 

At this stage, individual images can still contain gradients in the background, most likely resulting from the broad time interval that, by default, groups all data together to build the background model. As a consequence, sampling variations in the sky background across an observation sequence is suppressed. For this reason, the pipeline computes another heavily smoothed background model at this stage to correct for any remaining large-scale gradients visible in the frames.

We note here that it is also possible to construct and subtract a background model prior to flat-fielding. In this way, no correction for the dark current with separate calibration files is required. In our tests, however, we have found that the processed images deliver marginally better noise characteristics when applying the background correction only after flat-fielding.

\subsection{Weight maps}
\label{ssec:weightmaps}

Using the post-processed pawprints, the pipeline individually constructs weight maps, which are based on the main flat field for each image. Optionally, cosmic rays can be identified by Laplacian edge detection \citep{vanDokkum01,astroscrappy}. Furthermore, the pipeline also supports the MaxiMask algorithm for identifying contaminants (e.g., hot pixels, persistence, and fringe patterns) in astronomical images \citep{maximask}. For any identified contaminant, the corresponding pixels on the weight maps are set to 0. By default, these options are deactivated because they significantly increase the processing time for observation blocks and they can mask bright sources under excellent seeing conditions. Additionally, we did not notice significant quality gains in terms of noise properties in typical data products from the VISIONS survey when contaminant detection was enabled. This is most likely a consequence of the effective bad pixel rejection during stacking.

\subsection{Astrometric solution}
\label{ssec:astrometric_solution}

For the astrometric solution, we used Scamp \citep{scamp}, which requires source catalogs, generated by SExtractor \citep{sextractor}, as input. The pipeline first runs SExtractor on all post-processed pawprints (also pointing to their respective weight files) with a setup preset that is geared to typical VISIONS observations. Following source extraction, Scamp uses the object catalogs to perform cross-identification, to attempt pattern matching, and to compute an astrometric solution that describes the distortion pattern of the focal plane array with polynomials. For VISIONS, the pipeline includes a setup preset for Scamp that, by default, computes a distortion polynomial of fourth order. Moreover, we have tested several configurations of how Scamp treats individual FITS files and extensions and find that individually solving each exposure was the only setup that reliably delivered accurate astrometric solutions for VIRCAM. As a reference catalog, the 2MASS and \textit{Gaia} DR3 \citep[][]{GaiaDR3} databases can be queried, with the latter being the default option.

The pipeline also offers the possibility of providing externally computed astrometric solutions, making it possible to refine the calibration outside of the standard workflow. Specifically, for VISIONS we have built observation-specific reference catalogs based on \textit{Gaia} DR3 source positions. Given the proper motions measured with \textit{Gaia}, we transformed the positions from the \textit{Gaia} DR3 reference epoch (2016) to the epoch of the observations before computing an astrometric solution. Thus, the mean proper motion of the observed field has not been shifted to the epoch of the reference catalog, but instead it allows one to calibrate input data with respect to the epoch of the observations.

\subsection{Illumination correction and zero points}
\label{ssec:illumination_correction_and_zp}

Inspecting the post-processed images, we have found that, despite flat-fielding earlier, zero points at this stage can still diverge by up to a few tens of percent across the focal-plane array. To provide a constant, flat zero point for all provided frames, the pipeline constructs an illumination correction based on an external reference catalog. Since all VISIONS observations have been taken in the $J$, $H$, or $K_S$ bands, the 2MASS source catalog can serve as an external calibrator for all collected data. Moreover, using an external reference allows all data to be scaled to a single common zero point, regardless of any overlaps between images.

Following the computation of the astrometric solution, the image coordinates of the detected sources can be converted to accurate world coordinates. First, the pipeline runs SExtractor on all post-processed pawprints with a specific setup preset, also taking the currently available astrometric calibration into account. Given the sources' translated positions in right ascension and declination, the recipe performs a crossmatch with the 2MASS source catalog. A crossmatch between the cleaned 2MASS and VISIONS source catalogs is considered valid if the coordinates are within \SI{1}{arcsec} of each other. From the crossmatch, only sources with a 2MASS photometric quality flag equal to A (signal-to-noise ratio > 10) and an artifact contamination flag of zero (sources are not affected by known artifacts) are retained. At the same time, a quality filter based on the VISIONS source catalogs is applied. Here, we require sources to have a peak flux smaller than \SI{80}{\percent} of the given saturation limit, to have no detected neighbors within 5 pixels, to be located at least 10 pixels from the detector edges, and to have full width half maxima (FWHM) and ellipticities that are largely compatible with point sources (\SI{0.8}{pix}~<~FWHM~<~\SI{6.0}{pix} and ellipticity~<~0.25). Given these criteria, for each detector and depending on the local source density, several dozen to several hundred matches are available.

For all matched sources, the pipeline calculates the difference between instrumental and 2MASS catalog magnitudes. For each detector, the measured individual magnitude differences are then projected onto a grid with a size of $20 \times 20$ cells. For each cell in the grid, outliers are removed by sigma clipping and an average is calculated using the distance to the grid center positions and the inverse of its squared magnitude error as weights. This weighting method is used in multiple subsequent pipeline recipes to map a distribution of values, given their coordinates, to a grid that matches the input images and generally takes the form
\begin{align}
    \bar{x} & = \frac{\sum_i x_i \cdot w_i}{\sum w_i} \label{equ:weighted_average1} \\
    w_i & = \sigma^{-2}_{m_{\mathrm{tot}}} \cdot \exp{\left(-0.5 \cdot d^2 / s^2 \right)},  \label{equ:weighted_average2}
\end{align}
where $\bar{x}$ is the desired weighted average, $w_i$ are the weights, $\sigma_{m_{\mathrm{tot}}}$ are the total magnitude errors, $d$ is the distance between the object and the reference point, and $s$ is the standard deviation of the Gaussian kernel. The distance $d$ can be specified in image coordinates or as an on-sky separation in \si{arcmin}, for example. In the case of the illumination correction, the magnitude errors and the standard deviation correspond to $\sigma_{m_{\mathrm{tot}}}^2 = \sigma_{m_{\mathrm{2MASS}}}^2 + \sigma_{m_{\mathrm{instr.}}}^2$ and $s = 3\,\si{arcmin}$, respectively. The resulting grid is then further smoothed with a two-dimensional median filter with a three-cell-wide footprint and convolved with a two-cell-wide Gaussian kernel. Finally, the filtered grid is upscaled to match the original dimensions of the input images using bi-cubic interpolation.

\begin{figure}
        \centering
    \resizebox{\hsize}{!}{\includegraphics[]{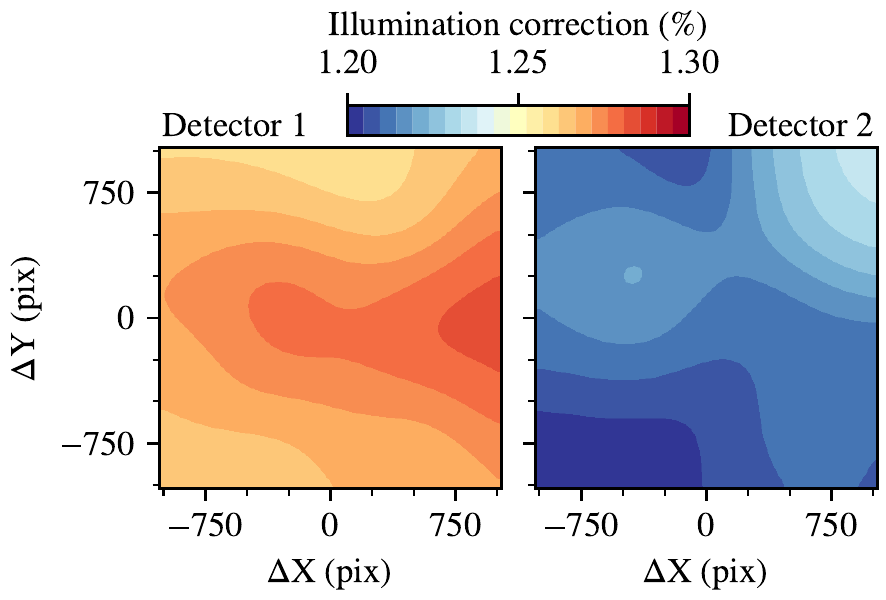}}
        \caption{Example for an illumination correction for the VIRCAM detectors 1 and 2. After flat-fielding, we found residual zero-point offsets up to \SI{10}{\percent} for individual pawprints. The VISIONS pipeline computes scaling factors for each image that equalize the zero points to a constant value of \SI{25}{mag}.}
        \label{img:ic}
\end{figure}

The procedure outlined above results in a heavily smoothed map that can be used to correct inconsistencies in the originally applied flat field. Specifically, for VISIONS, this illumination correction is further scaled so that the zero point of each image evaluates to $m_{\mathrm{zp}} = 25\,\si{mag}$. This has the advantage that the flux in the frames of any observation block in the entire VISIONS survey is scaled to the same zero point, and therefore any combination of images can be co-added at a later date without rescaling. Figure~\ref{img:ic} shows an example of the illumination correction for VIRCAM detectors 1 and 2, visualizing the residual zero-point offset not only between adjacent detectors, but also across individual images.

\subsection{Resampling and co-addition}
\label{ssec:resampling_and_co-addition}

\begin{figure*}
        \centering
    \resizebox{\hsize}{!}{\includegraphics[]{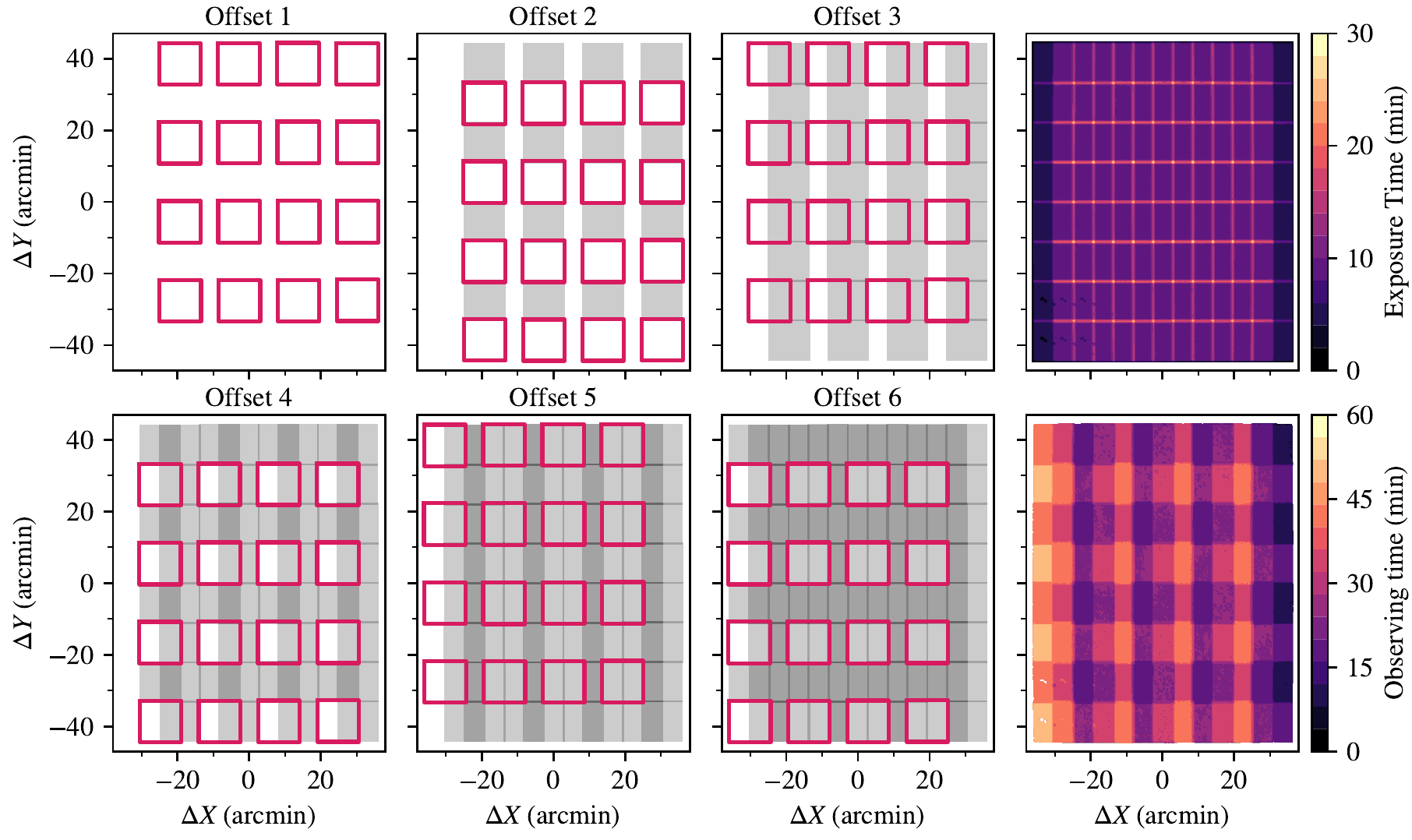}}
        \caption{Offset tile pattern "Tile6zz" and the resulting inhomogeneous sky coverage for co-added data products. The labeled panels show the consecutive construction of a tile from a series of six telescope offset positions. In each of these panels, the red boxes mark the current coverage of the VIRCAM pawprint, which is shown in gray in the subsequent offset steps. The two panels on the right show the effect of this observation strategy on the effective exposure time and the relative observing time for a typical observation block in the VISIONS program.}
        \label{img:tile_coverage}
\end{figure*}

Using the individual weight maps (Sect.~\ref{ssec:weightmaps}), the astrometric solution (Sect.~\ref{ssec:astrometric_solution}), and the illumination correction (Sect.~\ref{ssec:illumination_correction_and_zp}), all science frames can be rescaled and resampled onto a common reference projection. In particular, for VISIONS, these projections are predefined, so that all observed fields for a given star formation region share the same world and pixel reference coordinates. The advantage of this approach is that all individually resampled pawprints  (or any subset) for a specific region can be combined without performing another reprojection. For example, all individual frames for a given wide survey region can be combined into a mosaic, which includes the entire survey area in a single file. Together with the previously applied scale for matching the zero points, this choice allows one to perform source extraction and photometry for a region with a single task and without the need for subsequently combining source catalogs and dealing with overlapping areas. For all regions, we have constructed a zenithal equal-area (ZEA) projection in equatorial reference coordinates with a pixel scale fixed at a value of \SI{1/3}{arcsec\,pix^{-1}}. For all combinations, the pipeline uses the clipped co-addition method developed by \citet{Gruen14} with default values for the allowed flux variations.

For the resampling procedure, SWarp is used with an identical setup configuration for all VISIONS data. Given the weight maps discussed in Sect.~\ref{ssec:weightmaps}, SWarp resamples the images with a third-order Lanczos kernel, without flux scaling or background subtraction, since these steps have already been performed in the course of previously applied recipes. Although SWarp has the capability of applying a photometric scaling factor based on internal crossmatches, this option is only applied to the focal plane array as a whole. As a consequence, SWarp cannot correct for a spatially variable zero point between or within the detectors. This limitation represents another motivation for applying the illumination correction and fixed zero-point scaling prior to resampling. Running SWarp results in a set of resampled images, which are astrometrically and photometrically calibrated and, therefore, science-ready. As indicated in Fig.~\ref{img:flowchart}, these files represent the basis for all data products, including stacks, tiles, mosaics, files that are compliant with the ESO phase 3 standard, and curated source catalogs.

Due to the nature of the observing sequence of a VIRCAM observation block, several parameters of interest are not homogeneously distributed across a co-added image product. Figure~\ref{img:tile_coverage} visualizes this characteristic by showing the offset sequence and its effect on the data products. The displayed sequence corresponds to the "Tile6zz" offset pattern which was used for all observations in the VISIONS survey. The panels with the numbered labels show the six offset positions of the 16 detectors in a VIRCAM pawprint highlighted as red squares. For each subsequent step, the telescope executed predefined offsets, eventually covering the entire tile, where the gradual filling is shown in gray. The two panels on the right-hand side display the effect of the offset sequence on the effective exposure time and the relative observing time. The parameters in question appear in the style of a checkerboard pattern as a consequence of the overlaps between the detectors throughout an observing sequence. The effective exposure time in a tile resembles the coverage of the offset pattern, mixed in with individually rejected bad pixels, as it is visible in the bottom left part in the corresponding panel. The effective observation time displays an even more complex pattern, where again a checkerboard arrangement becomes visible; this parameter, however, also includes an overall gradient across the entire tile.

To extract the correct information for each source, the pipeline generates scaled-down versions of the input pawprints in the form of additional image files containing information on the number of images taken at a given position, the exposure time, the Julian date of the observation, and the estimated astrometric error (as calculated by Scamp). These files are then co-added with SWarp to create an image that has the same shape as the tile (or stack) created from the post-processed pawprints and are later used to feed information of these spatially variable parameters to the source catalogs. 

From the resampled pawprints, a series of data products can be created. In the case of the VISIONS survey, the pipeline assembles three different types of co-added images. Stacks refer to co-added jitter sequences at a given offset position. Since the observing strategy with VIRCAM requires a total of six pawprints to form a contiguous tile, every observation block produces six stacks. Tiles combine all pawprints of one observation block into a single co-added image. Finally, we refer to a co-addition of an arbitrary number of resampled pawprints as mosaics. An example of such a product in the VISIONS survey would be all the data from the wide survey for a particular star formation region. While stacks and tiles are submitted to ESO as part of the Phase 3 commitment, mosaics are built to create curated source catalogs that we plan to publish via the data repositories managed by the Centre de Données astronomiques de Strasbourg\footnote{\href{https://cdsweb.u-strasbg.fr}{https://cdsweb.u-strasbg.fr}} (CDS).

\subsection{Source catalogs}
\label{ssec:source_catalogs}

As visualized in Fig.~\ref{img:flowchart}, the resampled pawprints can be combined into three separate image data products. Depending on each individual case, the pipeline applies a specific series of recipes for source extraction and catalog post-processing.

\subsubsection{Source extraction}
\label{sssec:source_extraction}

Source extraction is performed with SExtractor in all cases. Similarly to source extraction for the astrometric calibration, the utilized setup preset includes the configuration for SExtractor as well as a list of parameters that need to be measured. For the source extraction, the detection threshold is set to $1.5\,\sigma$, while requiring a cluster of at least three connected significant pixels. For source detection, the images are convolved with a Gaussian kernel with a standard deviation of \SI{2.5}{pix} ($\approx 0.83\,\si{arcsec}$). The parameters that are extracted include, among others, windowed positions in image and world coordinates and their corresponding errors, shape profiles (e.g., FWHM, ellipticity, and isophotal area), flags, signal-to-noise ratio, as well as fluxes and magnitudes measured with both adaptively scaled and fixed-size apertures. Regarding the apertures with fixed sizes, in total 16 measurements were performed for each source, with aperture diameters ranging from three to 30 pixels, which, given the fixed plate scale of \SI{1/3}{arcsec\,pix^{-1}}, correspond to 1 to \SI{10}{arcsec}.

\subsubsection{Source catalog post-processing}
\label{sssec:catalog_post-processing}

The source tables generated by SExtractor are further processed in a manner that includes the addition of several observation-specific parameters. Although the pipeline has, at this point, already scaled the flux in all image products to a fixed value of \SI{25}{mag}, the workflow includes two more calibration steps for the measured fluxes. First, aperture matching is performed for all fixed-size apertures before the zero points are reevaluated and adjusted.

For aperture-matching, the recipe calculates the difference in measured flux with respect the largest aperture (i.e., \SI{10}{arcsec}) for each source in the SExtractor catalogs and for each aperture. Next, a sub-sample is constructed from the input catalog. It includes only point-like sources that do not have a detected neighbor within \SI{2}{arcsec}, which could have contaminated the flux measurement. Here it should be noted that it would have been desirable to reject all sources with neighbors within the radius of the largest aperture. However, with this criterion, most sources would have been filtered. Next, the aperture correction is estimated by calculating the weighted average at its position against the cleaned sub-sample for all detected sources. Similarly to the estimate of the illumination correction, the position-dependent, weighted, average value is computed following Equs.~\ref{equ:weighted_average1} and \ref{equ:weighted_average2}. Now, however, the standard deviation of the Gaussian component of the weight function is set to \SI{3}{arcmin} and $\sigma_{m_{\mathrm{tot}}}$ are the magnitude errors measured by SExtractor. This procedure independently derives values for the aperture correction for each source, based on a locally evaluated shape of the point spread function (PSF). We would like to note here that the measured aperture correction is only valid for point sources and specifically for bright, spatially resolved galaxies for which small fixed-size apertures do not represent their total flux well. 

Due to earlier rescaling, all images already feature a fixed zero point of \SI{25}{mag} prior to reprojection and co-addition. However, for source catalogs generated from co-added pawprints, we typically measure residual variations of the zero point of \SI{1}{\percent} within a given image. These variations are likely a result of the resampling procedure where also differences in the measured flux, originating in the variable pixel scale in a pawprint, are taken into account. To correct for any residual bias in the source catalog magnitudes, the pipeline computes and applies -- for each source individually and analogously to the aperture matching procedure -- a spatially variable correction factor to the zero points. Zero-point errors are calculated as standard errors for the sample of sources that matches the 2MASS source catalog. With the steps described above, the final calibrated catalog source magnitudes are then given by
\begin{equation}
    m_\mathrm{final} = m_{\mathrm{instr.}} + m_{\mathrm{apc}} + m_{\mathrm{zp}} + m_{\mathrm{zpc}},
\end{equation}
where $m_{\mathrm{instr.}} = -2.5 \log F$ are the instrumental magnitudes measured, with $F$ as the flux measured in ADU and $m_{\mathrm{apc}}$ and $m_{\mathrm{zpc}}$ are the smoothed, spatially variable aperture and zero-point corrections, respectively. Furthermore, $m_{\mathrm{zp}}$ denotes the fixed zero point of \SI{25}{mag} of all processed images, resulting from the illumination correction stage earlier in the workflow. 

The magnitude errors are estimated by SExtractor via
\begin{equation}
\label{equ:magerr}
    \sigma_{m,\mathrm{instr.}} = \frac{2.5}{\ln10} \frac{\sqrt{A\sigma^2 + F/g}}{F},
\end{equation}
where $A$ is the area that is used to measure the total flux $F$, $\sigma$ is the standard deviation of noise estimated from the background, and $g$ denotes the detector gain. This equation does not include any information on noise introduced by the flux calibration, aperture matching, or residual flat-field errors. Hence, the magnitude errors provided by SExtractor are underestimated in all cases and we refer to $\sigma_{m,\mathrm{instr.}}$ as instrumental errors. These underestimated values are specifically prominent for bright sources, where the error budget is dominated by the measured total flux. To obtain a more realistic magnitude error, we have compared flux measurements of the sources that were imaged in multiple epochs of the VISIONS survey. Details on this analysis are given in Sect.~\ref{ssec:performance_photometry} where we evaluate the pipeline performance with respect to photometric measurements. For bright sources, we find that the distribution of measured magnitudes across different observing epochs is well described by a Gaussian with a standard deviation of about \SI{0.005}{mag}. To obtain the final magnitude errors, this value was added to the errors determined by SExtractor
quadratically,\begin{equation}
\label{equ:magerr_final}
    \sigma_{m,\mathrm{final}} = \sqrt{\sigma_{m,\mathrm{instr.}}^2 + 0.005^2},
\end{equation}
thus effectively setting an error floor of \SI{5}{mmag} for the VISIONS survey. The value of the photometric noise floor will be reevaluated for every data release in the future.

In addition to providing aperture-matched magnitudes, the VISIONS pipeline optionally constructs an optimized magnitude $(m_\mathrm{best})$ that is based on aperture-matched fluxes in the 16 fixed-size apertures. To this end, the recipe compares the isophotal area of each source, as measured by SExtractor, to the available aperture sizes and chooses the best match. This workflow step selects apertures with sizes of 3 or \SI{4}{pix} for about two-thirds of all extracted sources.

The final step in the post-processing module for source catalogs adds information on the number of images at a given position, the exposure time, the observing date, and an additional systematic error component. These data are taken from the additional files produced during co-addition (see Sect.~\ref{ssec:resampling_and_co-addition}) by extracting the corresponding pixel value at any given world coordinate position of a source.

\subsubsection{Source classification}
\label{sssec:source_classification}

\begin{figure}
        \centering
        \resizebox{\hsize}{!}{\includegraphics[]{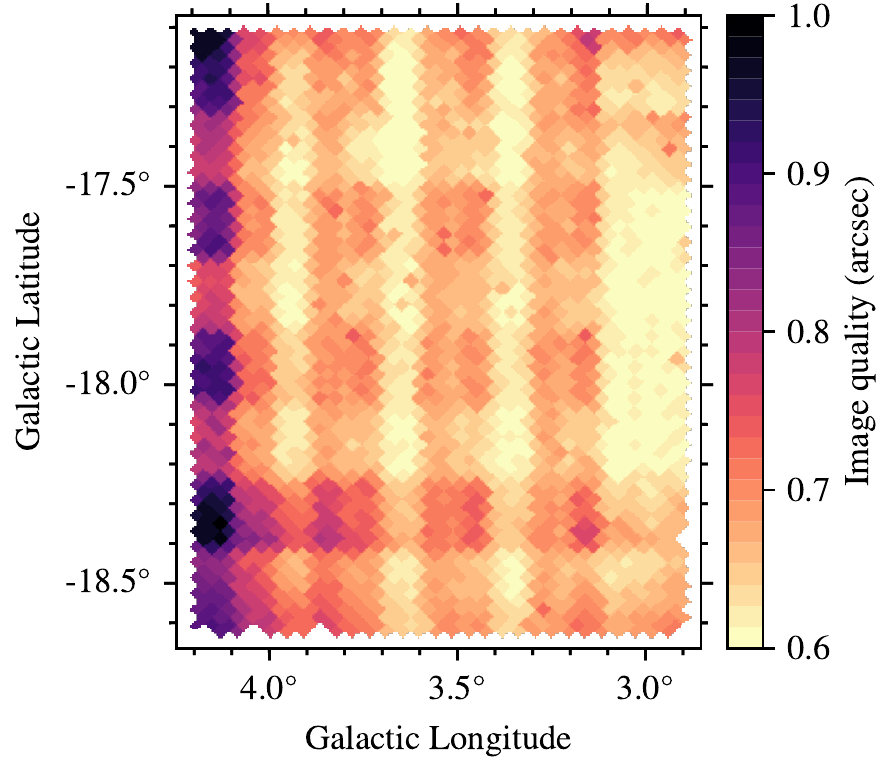}}
        \caption{Stellar FWHM measured for the Corona Australis control field in the $H$ band, mapped onto an 11th order healpix grid (pixel size = \SI{1.7}{\arcmin}). The observation offset sequence of the telescope creates a checkerboard pattern in the image quality, rendering source classification particularly complex.}
        \label{img:fwhm}
\end{figure}

The purpose of the source classification recipe is to discriminate between point sources and extended objects, with the two sets mainly comprising stars and galaxies, respectively. The source classification implemented in the VISIONS pipeline is based on the functionality built into SExtractor that uses a multilayered neural network to discriminate between extended (fuzzy) and point-like morphology. SExtractor generates a number between zero (extended) and one (point-like) as an indicator of how confidently objects were classified. The classifier has been trained on a set of 600 synthetic images created with SkyMaker \citep{skymaker}, resulting in a weight file that can be directly used during source extraction with SExtractor. This method has been shown to work robustly \citep{Philip02} and its previous application to VIRCAM data of the Orion star formation region \citep{meingast16} has delivered reliable results as well. 

One disadvantage of the classifier integrated in the SExtractor application is that a user is required to specify the expected FWHM for point sources as a fixed prior. In the case of VIRCAM observations, however, this parameter strongly varies across a co-added tile. The highly variable stellar FWHM for VIRCAM is a consequence of the large field of view, which results in the outer detectors typically having \SI{10}{\percent} larger PSFs than their central counterparts. A complex pattern in image quality becomes apparent across a tile as a consequence of the co-addition of at least 12 (in the case of the VISIONS wide survey, 30) input pawprints distributed over sometimes tens of minutes long observation blocks. This effect is visualized in Fig.~\ref{img:fwhm}, where the image quality, parametrized with the FWHM of bright point sources, is displayed on a colorscale for the VISIONS $H$-band control field observations of the Corona Australis star-forming region. The figure highlights a checkerboard pattern which is a result of variable seeing and the specific telescope offset pattern during the execution of the observation block. 

Due to the highly variable PSF across all images, SExtractor is not able to deliver robust results regarding object classification in a single execution. To derive a trustworthy discrimination between sources with an extended or point-like morphology, the VISIONS pipeline first estimates the absolute range in stellar FWHM for a given source catalog. Subsequently, a lookup table is constructed by extracting and classifying sources multiple times with variable input seeing values in steps of \SI{0.05}{arcsec}. Next, the pipeline creates a two-dimensional distribution of stellar PSF sizes similar to previous interpolation methods (Equs.~\ref{equ:weighted_average1} and \ref{equ:weighted_average2}). Drawing upon the estimate of the local stellar FWHM from the smoothed distribution (as shown in Fig.~\ref{img:fwhm}), the pipeline obtains the final classification by interpolating the values in the previously built lookup table.

\subsubsection{Quality flags}
\label{sssec:quality_flags}

Inspired by the 2MASS source catalog, the pipeline constructs several subsets from the post-processed source catalogs, which correspond to different quality criteria. This includes the extraction flags determined by the SExtractor FLaG (SFLG), a Contamination FLaG (CFLG), and a Quality FLaG (QFLG). SExtractor flags comprise eight bits that indicate issues with flux measurements (e.g., neighboring sources and saturated pixels), morphological characteristics (e.g., truncated footprint), and memory overflows, which can result in incomplete flux measurements. The definition of these flags is available through the documentation of the tool\footnote{\href{https://sextractor.readthedocs.io/en/latest/Flagging.html}{https://sextractor.readthedocs.io/en/latest/Flagging.html}}. The boolean contamination flag combines a set of criteria to filter unreliable source detections, including objects with negative SExractor flux measurements, a FWHM smaller than \SI{0.2}{arcsec}, sources that do not have a valid SExtractor source classification (Sect.~\ref{sssec:source_classification}), and objects with truncated PSFs. Additionally, the curve of growth constructed from the flux measurements in fixed-size apertures is examined for each entry in the source catalog. Sources are flagged as contaminants if their curve of growth shows a smaller increase in the measured flux than expected for point sources. This criterion is particularly effective for identifying detections of extended emission, which are exceptionally numerous in the vicinity of star-forming regions. The quality flag QFLG has been designed analogously to the 2MASS source catalog. This flag can take the values A, B, C, D, or X given the following criteria:

\begin{itemize}
    \item A: $(\mathrm{SFLG} < 4) \cap (\mathrm{CFLG} = 0) \cap (\sigma_{m,\mathrm{best}} < 0.10857\,\si{mag})$;
    \item B: $(\mathrm{SFLG} < 4) \cap (\mathrm{CFLG} = 0) \cap (\sigma_{m,\mathrm{best}} < 0.15510\,\si{mag})$;
    \item C: $(\mathrm{SFLG} < 4) \cap (\mathrm{CFLG} = 0) \cap (\sigma_{m,\mathrm{best}} < 0.21714\,\si{mag})$; and
    \item D: $(\mathrm{SFLG} < 4) \cap (\mathrm{CFLG} = 0)$.
\end{itemize}
All remaining sources are assigned the flag X. The criterion $\mathrm{SFLG} < 4$ keeps deblended sources with potentially biased aperture photometry, but filters incomplete isophotal footprints and truncated sources. In general, we recommend working with the subset where $\mathrm{QFLG} \neq X$ and, depending on each individual case, also enabling further restrictions in the quality flag to retain only sources with high signal-to-noise ratios. Another very efficient method of filtering contaminants is either a crossmatch in multiple bands for the deep and control surveys, or requiring a detection in multiple epochs for the wide survey.

\section{Pipeline performance}
\label{sec:performance}

In this section, we evaluate the performance of the pipeline with respect to astrometric and photometric properties. In addition, we assess the reliability of the source classification recipe, determine runtimes for typical VISIONS observation blocks, and present a direct comparison to data products delivered with the CASU processing infrastructure.

\subsection{Astrometry}
\label{ssec:performance_astrometry}

\begin{figure}
        \centering
    \resizebox{\hsize}{!}{\includegraphics[]{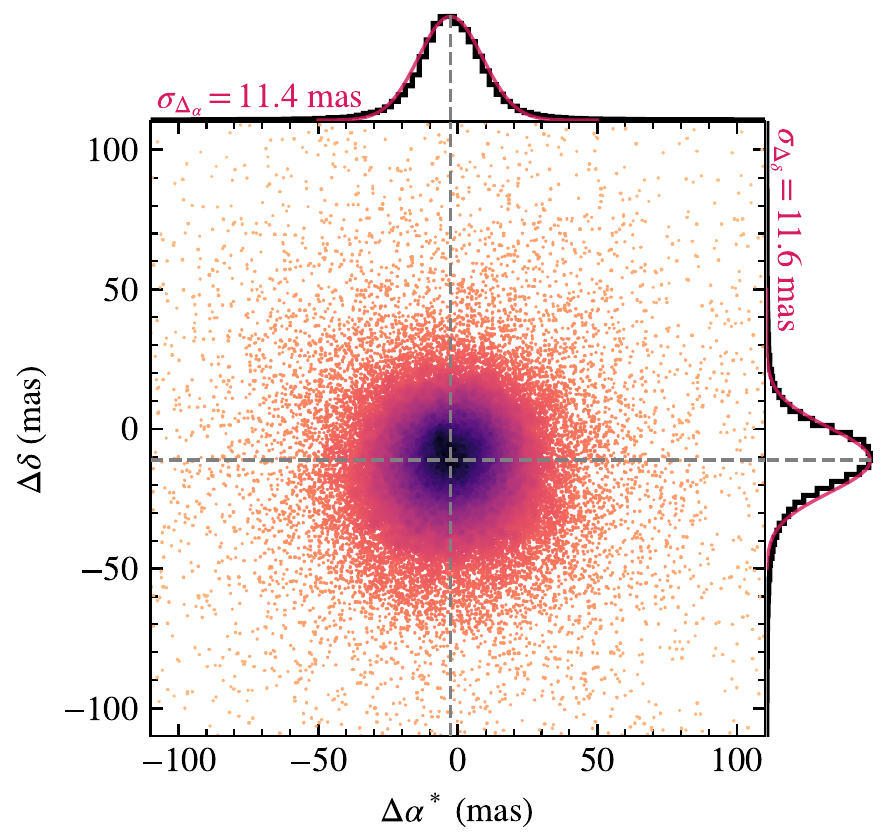}}
        \caption{Comparison of source coordinates determined by VISIONS with the \textit{Gaia} DR3 catalog for the Corona Australis control field. The y and x axes show the difference in right ascension and declination between \textit{Gaia} and VISIONS, respectively. All astrometric solutions for VISIONS were computed against a \textit{Gaia} frame that was shifted to the epoch of the individual observations. As a result, the distribution is not centered at the origin but coincides with the expected mean shift, marked by dashed gray lines between the two reference epochs (2016.0 for \textit{Gaia} DR3 and 2018.36 for VISIONS). The black histograms on the top and on the right-hand side of the plot are well described by a Gaussian (shown in red) with a mean at the expected shift and a standard deviation corresponding to the astrometric error reported by Scamp.}
        \label{img:astr_delta}
\end{figure}

To assess the total error budget in the VISIONS coordinates, we have compared the measured source positions with the subset of available matched source positions in the \textit{Gaia} DR3 catalog. The median uncertainty in the positions for \textit{Gaia} sources near the sensitivity limit of the survey at $G=21\,\si{mag}$ is \SI{1}{mas}; for brighter sources, the uncertainty decreases to tens of \si{\micro as} \citep{Lindegren21}. For the total error budget, proper motion measurements need to be considered as well since the pipeline transforms the \textit{Gaia} reference frame to the observing epoch. The errors in the proper motions are on the order of \SI{1}{mas\,yr^{-1}} for faint sources and of a few tens of \si{\micro as\,yr^{-1}} for bright sources. Given these typical values for the \textit{Gaia} position uncertainties, we have determined that the errors in the VISIONS source coordinates are predominantly governed by the statistical uncertainties arising from source extraction and any residual systematic errors originating from the astrometric solution. 

Figure~\ref{img:astr_delta} displays a comparison of VISIONS and \textit{Gaia} coordinates for the $H$-band measurements of the Corona Australis control field which was observed in May 2018. The figure shows the difference between the VISIONS and \textit{Gaia} source coordinates in right ascension and declination for a subset of measurements with a signal-to-noise ratio greater than 50 and a point-like morphology. The graphs on the top and on the right of the plot display the distributions as histograms and the color scale is proportional to the density of points in this parameter space. Since \textit{Gaia} coordinates are given for epoch 2016.0, the distribution in the figure shows an offset of about \SI{10}{mas} in declination and about \SI{2}{mas} in right ascension. This offset coincides with the mean motion of the sources in this field over the elapsed time of about \SI{2.4}{yr}, which we have determined to be $\bar{\mu}_{\alpha^*}=-1.1\,\si{mas\,yr^{-1}}$ and $\bar{\mu}_\delta=-4.73\,\si{mas\,yr^{-1}}$. The mean motion for the elapsed time is marked with dashed gray lines in the plot. The red lines overlaying the histograms represent Gaussian functions with a mean set to the expected shift between the observation epochs. The standard deviation is equal to the astrometric error reported by Scamp (\SI{11.7}{mas} for this field). The peak in the histograms is well matched to the mean motion of the field. Also, the distribution of sources in this diagram appears to be Gaussian-shaped, with a width dictated by the errors reported by the astrometric solution. These findings indicate that the coordinates determined by the VISIONS pipeline are well behaved and the absolute position errors reported in the source catalogs are reliable. For additional insights regarding the astrometric properties, we refer the reader to the study of \citet{Libralato15} who developed an independent reduction pipeline for VIRCAM, mainly focusing on high-precision astrometry.

\subsection{Photometry}
\label{ssec:performance_photometry}

\begin{figure}
        \centering
    \resizebox{\hsize}{!}{\includegraphics[]{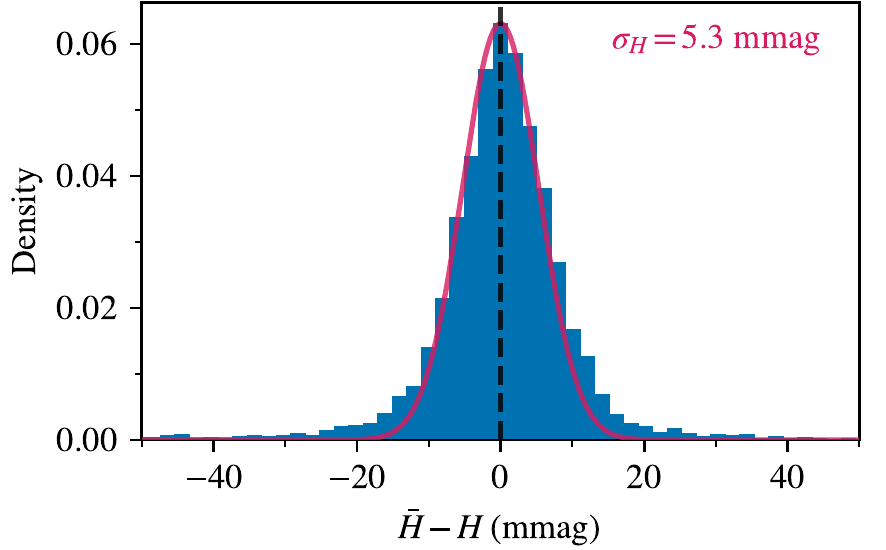}}
        \caption{Comparison of measured $H$-band magnitudes for all epochs of the VISIONS wide survey in Corona Australis. The blue histogram shows the difference of individually measured source magnitudes to their mean magnitude for bright sources ($12 < H < 13\,\si{mag}$). The red line depicts a Gaussian centered at zero, with a width determined by the median standard deviation of all included sources.}
        \label{img:photerr}
\end{figure}

To assess the pipeline's capabilities with respect to the flux calibration, we performed two separate tests. First, we compared the measured magnitudes of sources that were imaged over multiple epochs in the VISIONS wide survey. Second, due to the distinct VIRCAM and 2MASS filter systems, we explored whether any color terms remain in the output catalogs.

The magnitude errors computed by SExtractor (Equ.~\ref{equ:magerr}) do not include any systematic component that may have been introduced during the calculation of the zero point and aperture matching. In order to assess the noise floor of our photometric calibration, we compared source magnitudes from processed data across the six epochs of the wide survey in the Corona Australis star-forming region. First, we combined the catalogs by crossmatching sources with a maximum separation of \SI{1}{arcsec}. Subsequently, all matched sources were filtered, requiring $\mathrm{QFLG} \neq X$ and $\sigma_{m,\mathrm{instr.}} < 0.01\,\si{mag}$. Figure~\ref{img:photerr} displays a histogram of the difference between each measurement and the mean value in the six observing epochs for all remaining sources. The red line represents a Gaussian function centered at the origin, with a standard deviation given by the median standard deviation in the measured magnitudes for all observations corresponding to $\sigma_H = 5.3\,\si{mmag}$. The Gaussian is well correlated with the distribution, indicating that the latter is well represented by individual magnitude errors. Only the wings of the histogram diverge slightly from a Gaussian shape, which may be a result of intrinsic source variability. Given this analysis, we conclude that the noise floor in the photometry is at the millimag level, motivating the addition of this systematic term in the calculation of the final magnitude errors (Equ.~\ref{equ:magerr_final}).

\begin{figure}
        \centering
    \resizebox{\hsize}{!}{\includegraphics[]{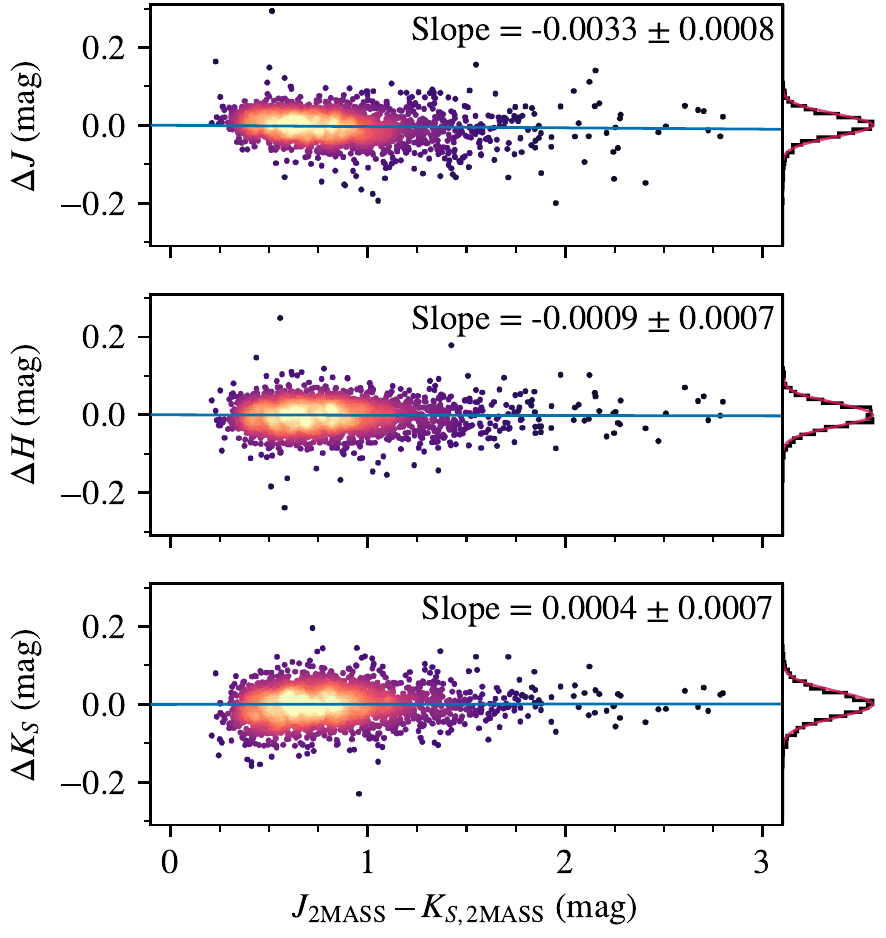}}
        \caption{Evaluation of residual color terms in the VISIONS photometry. The panels show the difference between VISIONS and 2MASS source magnitudes as a function of their $J-K_S$ color for the $J$ (top), $H$ (center), and $K_S$ (bottom) bands. The blue line indicates a linear fit, with the slope reported in the top right of each panel. The data are projected into a histogram (black) on the right-hand side. The red line on top of the histograms shows a Gaussian determined by the mean measurement errors in the source catalogs.}
        \label{img:color_transformation}
\end{figure}

In the second performance test for the VISIONS photometry, we evaluated the existence of residual color terms arising from the different filter systems of VIRCAM and 2MASS. To cover highly reddened sources as well, we have constructed a matched source catalog between the two surveys from the observations of the Corona Australis deep field. In addition, we required all sources to have a clean VISIONS quality flag ($\mathrm{QFLG} \neq X$), 2MASS magnitude errors smaller than \SI{5}{\percent}, a source classification more likely to indicate a point source than an extended one in all bands (CLS > 0.5), and no detected neighbors within \SI{3}{arcsec}. In Fig.~\ref{img:color_transformation} we show the differences between the 2MASS and VISIONS magnitudes in $J$, $H$, and $K_S$ as a function of the $J-K_S$ color of a source in 2MASS. The colormaps in all panels indicate source density, and the blue line represents a linear fit. The slope of the linear fit is annotated on the top right of each panel, also including an error determined by \SI{1000} bootstrap iterations. On the right-hand side of each panel, the data are displayed as a histogram (in black) that is overlayed by a Gaussian function, whose variance corresponds to the total error budget $\sigma_{m_{\mathrm{final}}}^2 + \sigma_{m_{\mathrm{2MASS}}}^2$.

The comparison between 2MASS and VISIONS source magnitudes as a function of color in Fig.~\ref{img:color_transformation} does not reveal noteworthy color terms for any passband. Only in the $J$ band do we find a slope in the magnitude-color relation that is statistically compatible with nonzero values. Furthermore, the scatter in the magnitude differences appears to be well described by a Gaussian determined by the total measurement error in both surveys. Given the fact that the distribution can be described by measurement errors only, combined with the sparse availability of highly reddened sources, and a slope measured at the millimag level, we conclude that no color transformation is necessary between the VISIONS and 2MASS photometry. Nevertheless, this relation will be tested for all future data releases.

\subsection{Classification}
\label{ssec:performance_classification}

\begin{figure}
        \centering
        \resizebox{\hsize}{!}{\includegraphics[]{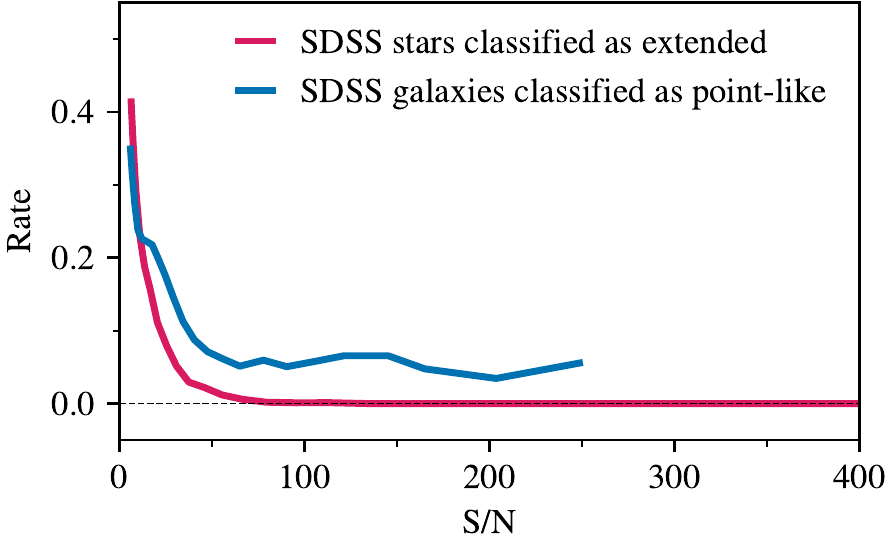}}
        \caption{VISIONS source classification compared to the one found in SDSS for a randomly selected tile in the VIKING survey. The red line shows the fraction of sources at a given signal-to-noise ratio, which are classified as stars in SDSS DR16 but were assigned an extended morphology in the VISIONS pipeline. Similarly, the blue line depicts the fraction of SDSS galaxies that were identified as being point-like in VISIONS. For both stars and galaxies, the rate of misclassifications rises for sources with low signal-to-noise ratios. In addition, for SDSS galaxies we find a constant threshold at about \SI{5}{\percent} as a consequence of insufficient spatial resolution of the VISIONS survey.}
        \label{img:class_sdss}
\end{figure}

To estimate the performance of the source classifier in the VISIONS pipeline, we compared VIRCAM data to the star-galaxy discrimination as published in data release 16 of the Sloan Digital Sky Survey \citep[SDSS;][]{sdss_dr16}. To this end, we have obtained raw VIRCAM data from a random field in the VISTA public survey VIKING \citep{viking} that has also been imaged in SDSS (observation block ID 1121101). For the test, we processed the $J$-band VIRCAM observations of this field with the VISIONS pipeline to obtain the classification value, as outlined in Sect.~\ref{sssec:source_classification}. Figure~\ref{img:class_sdss} shows a comparison between the results of the pipeline classification and the object type determined in the SDSS data. The solid red line in the diagram shows the fraction of sources that are classified as stars in SDSS, but are assigned an extended morphology (CLS < 0.5) in VISIONS as a function of the signal-to-noise ratio. Only at relatively low signal-to-noise ratios does the pipeline classification scheme start to fail, and point sources are increasingly identified as extended objects. Similarly, the solid blue line shows galaxies, as determined by SDSS, that were assigned a point-like morphology in VISIONS. Also, the misclassification rate increases for low signal-to-noise ratios. Furthermore, we observe a constant rate of misclassification of about \SI{5}{\percent}, independent of the brightness of the source. An examination of the images and source colors of the objects in question in the SDSS catalog reveals that they are indeed likely galaxies, but that VISIONS could not spatially resolve them.

Overall, the approach of discriminating between an extended and point-like morphology in the VISIONS pipeline delivers exceptional results, with only a fraction all sources being misclassified. However, the performance of the classifier largely depends on the image quality, since the number of spatially resolved galaxies decreases with degrading seeing conditions during the observations. In the case of the selected VIKING field, the threshold of misclassified galaxies increases to about \SI{10}{\percent} for the $K_S$ band, where the median image quality (FWHM) is about \SI{0.8}{arcsec}, compared to \SI{0.65}{arcsec} in the $J$ band.

\subsection{Benchmark}
\label{ssec:benchmark}

\begin{figure}
        \centering
        \resizebox{\hsize}{!}{\includegraphics[]{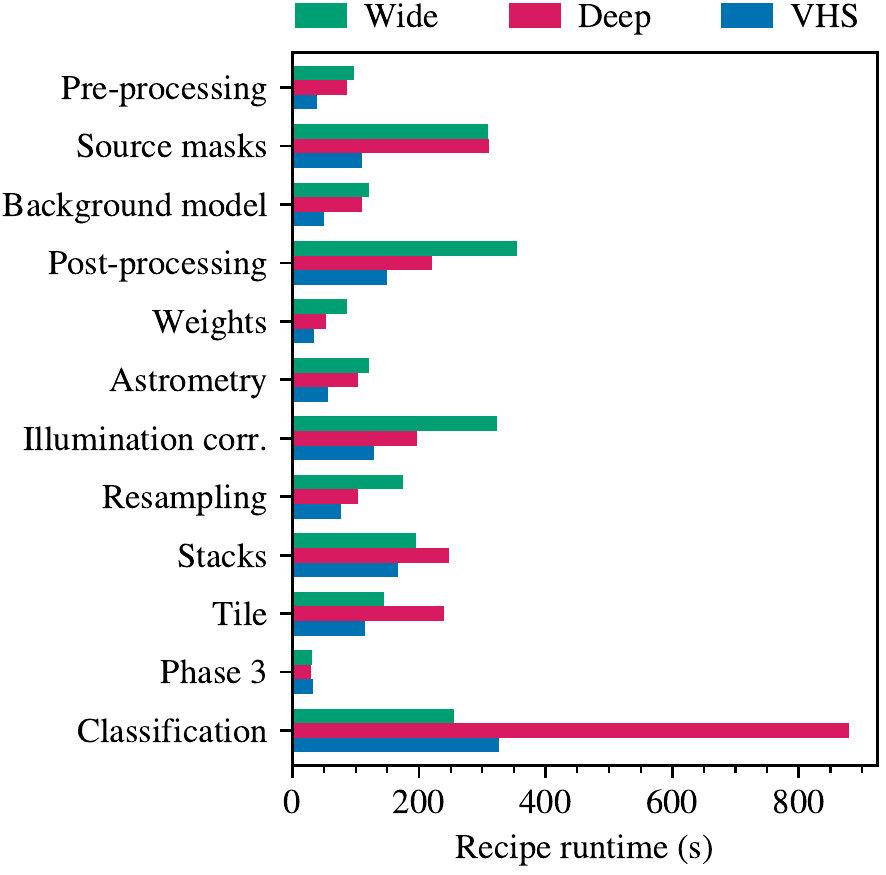}}
        \caption{Pipeline performance benchmark for three typical observation blocks in the VISIONS program. The horizontal bars show the recipe runtimes for a set of wide data in green, deep data in red, and a VHS sequence in blue. The processing time depends on the number of observed pawprints and the number of extracted sources.}
        \label{img:benchmark}
\end{figure}

To assess the performance of the pipeline in the version implemented at the time of writing, we have processed three observation blocks that are representative of typical VISIONS observations. The data included are sets of observations in the wide and deep subsurvey as well as a sequence recorded in the VISTA Hemisphere Survey \citep[VHS;][]{vhs}. The wide observation block has been observed as part of the Corona Australis region and includes 30 pawprints comprising five jittered positions at each of the six telescope offsets. The deep data correspond to a single observation block of the $J$-band tile of the Corona Australis complex. This set includes 18 pawprints on the target, as well as nine additional sky offset pointings. The data selected to profile the VHS tile is associated with observation block ID 557847 and includes 12 pawprints. All results have been obtained by running a single pipeline instance on an Apple MacBook Pro (M1 Pro CPU) with all data stored on the internal SSD and using Python 3.11.0 and NumPy 1.23.4.\footnote{We do not mention additional pipeline dependencies because the overall performance mostly depends on the Python version and the NumPy implementation.}.

To create a clearer benchmark profile, we have grouped the pipeline recipes into categories that follow the simplified data flow, as shown in Fig.~\ref{img:flowchart} and on the categorical axis of Fig.~\ref{img:benchmark}. Some recipes are not included in our benchmark, such as quality control plots and determinations of image statistics. However, these only contribute to a minor fraction of the overall execution time and do not significantly impact the processing speed. Figure~\ref{img:benchmark} shows the recipe runtimes for the three observed sets. In this figure, the green, red, and blue horizontal bars refer to data from the wide, deep, and VHS data sets, respectively. The total runtimes (not visible in the figure) are \SI{2320}{\second} for the wide set, \SI{2660}{\second} for deep data, and \SI{1347}{\second} for the VHS observation block. The pipeline modules, from pre-processing to resampling, show an approximately linear performance dependency on the number of input pawprints, where we observe the fastest processing speeds for the VHS set with only 12 pawprints. The wide and deep data sets follow this trend. However, some differences, which we attribute to the sky offset observations, become evident. For instance, we observe a similar performance for these two sets regarding the construction of source masks, but shorter runtimes for deep data during post-processing. While the former recipe runs on all pawprints in the observation block, post-processing includes only observations on the target and omits sky offset observations. As a result, the pipeline is oftentimes either I/O-bound, or the underlying function calls have running times proportional to the input size with $\mathcal{O}(n)$.

Unlike the execution times associated with array operations, source extraction has the largest impact on pipeline performance. This is evident in the classification runtimes visible in Fig.~\ref{img:benchmark}, where the data associated with the wide survey show the shortest execution times, while the deep data take more than twice as long. Here, the runtime increases approximately linearly with the number of sources in the image data products. Furthermore, specifically for the classification, the execution time can diverge further due to its dependency on the range of the measured FWHM of point sources (see Sect.~\ref{sssec:source_classification}). 

We note here that most tasks are I/O-bound, and the amount of data that are written to or read from the disk is highly variable throughout the processing of an observation block. As a consequence, depending on the available computation power and disk or network speeds, several pipeline instances can be run at the same time without a significant performance penalty.

\begin{figure*}
        \centering
    \resizebox{\hsize}{!}{\includegraphics[]{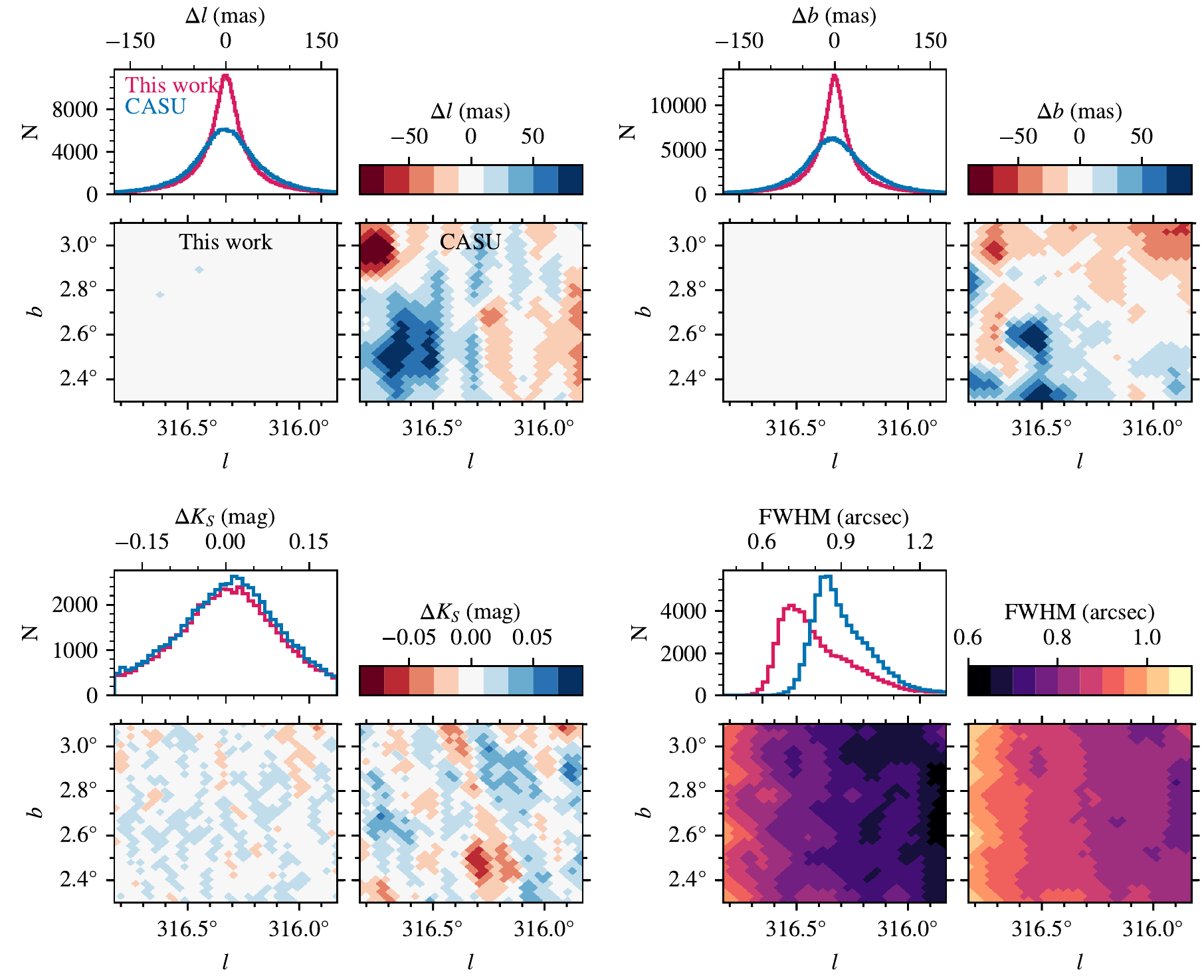}}
        \caption{Performance comparison between the VISIONS pipeline and the CASU infrastructure for a randomly selected tile in the VVV survey. The four distinct groups show histograms and spatial variability in the Galactic frame of coordinates relative to \textit{Gaia} (top panel groups), $K_S$-band photometry compared to 2MASS (bottom left), and the reported image quality parametrized by the stellar FWHM (bottom right). The red and blue histograms show the respective measurement distributions for the VISIONS and CASU pipelines. Concerning the spatial distributions, the panels on the bottom left and right-hand side show the properties derived by the VISIONS and CASU workflows.}
        \label{img:vcp_vs_casu}
\end{figure*}

\subsection{Comparison to CASU data products}
\label{ssec:performance_casu}

The VISIONS pipeline operates independently of any other data-processing facility of VISTA/VIRCAM. Given the effort of designing and implementing many recipes from scratch, we also tested the performance of the pipeline against publicly available data from the Variables in the Via Lactea Survey \citep[VVV;][]{vvv}. To this end, we downloaded a random VVV tile from the ESO archive (data observed in 2019 with observation block ID 2346480) and processed the files with the VISIONS pipeline. Subsequently, we compared the results from our pipeline with the ESO Phase 3 products of VVV that were processed with CASU pipeline version 1.5, with respect to their performance in astrometry, photometry, and image quality.

Figure~\ref{img:vcp_vs_casu} visualizes the comparison in four panel groups. In the top left and top right, differences in the measured Galactic longitude and latitude are displayed. The group on the bottom left-hand side compares the photometric consistency of both pipelines against 2MASS, and the panel group on the lower right displays the performance with respect to image quality. Each panel group consists of three subpanels, which include the respective metric in a histogram and two images (and a colorbar) that depict the spatial variations of the evaluated parameters. The panel on the bottom left of each group refers to the VISIONS pipeline, while the bottom panel on the right-hand side shows data processed with the CASU infrastructure, as submitted by the VVV consortium.

For the panel groups at the top of Fig.~\ref{img:vcp_vs_casu}, we compared the source positions reported in the VIRCAM source catalogs with the coordinates in the \textit{Gaia} database. For both longitude and latitude coordinates, the VISIONS pipeline does not show systematic uncertainties regarding the spatial distribution. On the contrary, the positions determined with the CASU pipeline show not only a significant but also a variable divergence from the reference position on the order of \SI{50}{mas} or more. In addition, the histograms in the top panel groups clearly show that the VISIONS pipeline outperforms other data products with respect to the precision in the measured coordinates.

The evaluation of the photometric performance, shown in the bottom left panel group of Fig.~\ref{img:vcp_vs_casu}, displays a comparison of the measured $K_S$-band magnitudes to their counterparts in the 2MASS source catalog. In this case and in contrast to the VISIONS pipeline, some spatial variability with amplitudes on the order of a few percent is also measurable for the CASU data products, but the histograms do not reveal significant differences. Furthermore, it should be noted that correction factors such as proposed by \citet{Hajdu20} can account for photometric offsets due to blended 2MASS sources. These corrections specifically, however, are most critical in areas of very high source density and therefore likely do not fully eliminate the variable offset displayed in Fig.~\ref{img:vcp_vs_casu}.

Given the width of the histogram, the comparison is likely limited by the measurement errors of the sources in the 2MASS source catalog. On the bottom right-hand side of Fig.~\ref{img:vcp_vs_casu}, we show the distribution of the VIRCAM image quality, parametrized by the FWHM of point sources with 2MASS quality flag A or B. Here, we also observe a stark difference between the two processing tools, with the VISIONS pipeline delivering about \SI{20}{\percent} smaller PSFs. As already explained by \citet{meingast16}, the degradation in image quality observed for the CASU pipeline is the result of the bilinear interpolation in the resampling algorithm. The spatial variation in both cases is a consequence of variable seeing conditions during the execution of the observation block.

\section{Pipeline data products}
\label{sec:data_products}

In general, the VISIONS pipeline produces flux-calibrated images and associated source catalogs. Regarding image data products, the pipeline produces VIRCAM stacks, tiles, and mosaics. Stacks refer to co-added pawprints that were recorded at one of the six offset positions and include multiple jittered observations. Tiles combine all data taken in the context of an observation block, comprising all offset and jitter positions. Mosaics refer to co-added images that are built from an arbitrary number of input pawprints (with an upper limit set by the available processing hardware). In the case of the VISIONS survey, this refers to the superposition of all observations within a single epoch and star-forming region. Furthermore, we discriminate between the data products submitted to ESO in the context of the Phase 3 commitment of the survey team and the results that are specifically geared for other forms of data dissemination.

\subsection{ESO Phase 3 compliance}
\label{ssec:phase3}

The survey team has committed to supply science-ready stacks and tiles, along with source catalogs, as part of ESO Phase 3. To this end, the corresponding data products from the VISIONS pipeline are further processed in dedicated recipes to comply with all Phase 3 requirements, as defined in the ESO Science Data Products Standard document\footnote{The Data Products Standard document is available at \href{https://www.eso.org/sci/observing/phase3/overview.html}{https://www.eso.org/sci/observing/phase3/overview.html}}. To facilitate compliance, the pipeline executes two separate recipes that modify the stacks and tile images as well as the catalogs. While the images themselves remain unchanged, several parameters are added or updated in the FITS headers, such as the keywords referring to the zero point and the setup for the observations (e.g., observation date). Here, the requirements dictate that we also supply the zero point in AB magnitudes. To this end, the pipeline transforms the measured Vega zero point with the equations given in \citet{Blanton05}. In addition, keywords for cross-identification and provenance are set. The contents of the source tables are taken from the results obtained during catalog post-processing (Sect.~\ref{sssec:catalog_post-processing}). We note here that not all recipes described in Sect.~\ref{sec:pipeline} are necessarily executed, and the contents of any data release will be described more thoroughly in the provided data release documentation.

\subsection{Curated source catalogs}
\label{ssec:curated_source_catalog}

Similarly to the procedure that produces ESO Phase 3 compliant images and source tables, the pipeline includes a stand-alone module that further modifies the data products resulting from all processing steps described in Sect.~\ref{sec:pipeline}. The purpose of this module is to produce a compact version of the source catalog that is more user-friendly than the data submitted to the ESO Phase 3 system. For instance, this module includes a recipe that integrates the 2MASS source catalog, so that saturated VISIONS sources are also included. Additionally, the catalog is purged of columns that are only used for catalog post-processing and thus do not represent an added value for the community. Using this strategy, a curated catalog similar in structure to the 2MASS survey can be constructed. The survey team plans to release these curated data products for each star-forming region via CDS.

\section{Summary}
\label{sec:summary}

In this paper, we have described the data-processing pipeline for the ESO VISIONS public survey. The pipeline is an end-to-end Python implementation (see data flow in Fig.~\ref{img:flowchart}) and processes raw VIRCAM data in the format in which it has been made available in the ESO science archive. The workflow includes the creation of main calibration files from data provided in the context of the VISTA calibration plan (Sect.~\ref{ssec:main_calibration}). These files were later used in a series of consecutively applied recipes (Sects.~\ref{ssec:pre-processing} through \ref{ssec:source_catalogs}) that process data collected on science targets. Said recipes removed the instrumental signature and provide science-ready images, for which the flux calibration was performed relative to the 2MASS source catalog. Astrometric solutions were derived relative to the \textit{Gaia} source catalog by shifting source positions to the observing epoch. Source catalogs have been built with SExtractor, and have been further processed to provide a range of measurements describing, for example, positions, magnitudes, and shapes for individual sources. 

We have evaluated the astrometric and photometric performance of the pipeline by comparing the results with the \textit{Gaia} and 2MASS reference catalogs. Due to the high precision of position measurements with \textit{Gaia}, the pipeline reaches an absolute astrometric accuracy on the order of \SI{10}{mas} for bright sources (Fig.~\ref{img:astr_delta}). Furthermore, the provided absolute photometry reaches a noise floor of approximately \SI{5}{mmag} (Fig.~\ref{img:photerr}), without significant color terms in any of the $J$, $H$, and $K_S$ bands (Fig.~\ref{img:color_transformation}). Moreover, the residuals with respect to both astrometry and photometry follow the expected Gaussian distributions, indicating reliable position and magnitude errors. 

Furthermore, we have assessed the capabilities of the pipeline by processing observations conducted as part of other VISTA surveys and comparing the results with publicly available data. In this comparison (Fig.~\ref{img:vcp_vs_casu}), we find that the VISIONS data-processing infrastructure provides superior results with respect to astrometry, photometry, and image quality. For data processed with other pipelines, we have found worrying evidence of spatially variable quality in position measurements, with offsets on the order of \SI{50}{mas}. Similarly, we have found a small spatially variable bias in the source magnitudes. However, we have not been able to determine the full extent because the comparison with 2MASS photometry is limited by the precision in the reference catalog. In all examined facets of the comparison between VISIONS and other, similar all-sky surveys or data-processing tools, our pipeline delivers source positions and magnitudes that are largely free from systematic errors.

The pipeline data products include images, source catalogs, and associated weight maps that are specifically prepared for the ESO Phase 3 publication process. In addition, the pipeline incorporates modules that generate curated source catalogs that are similar in structure to the 2MASS source catalog. Dissemination of the data products will be carried out within the ESO Phase 3 framework and curated image and source catalogs will be made publicly available through CDS.

\begin{acknowledgements}
We thank the anonymous referee for the useful comments that helped to improve this publication.
This research has used the services of the ESO Science Archive Facility.
This research has made use of Astropy\footnote{\href{http://www.astropy.org}{http://www.astropy.org}}, a community-developed core Python package for Astronomy \citep{astropyI, astropyII}.
This research has made use of "Aladin sky atlas" developed at CDS, Strasbourg Observatory, France \citep{aladin} and the table processing tools TOPCAT and STIL \citep{topcat}.
We also acknowledge the various Python packages that have been used for the preparation of this manuscript, including NumPy \citep{numpy}, SciPy \citep{scipy}, scikit-learn \citep{scikit-learn}, scikit-image \citep{scikit-image}, and Matplotlib \citep{matplotlib}.
This research has made use of the SIMBAD database operated at CDS, Strasbourg, France \citep{simbad}. This research has used the VizieR catalog access tool, CDS, Strasbourg, France (\citealp{vizier}; DOI: \href{http://dx.doi.org/10.26093/cds/vizier}{10.26093/cds/vizier}).
\end{acknowledgements}

\pagebreak
\bibliography{references.bib}

\end{document}